\documentclass[twocolumn,showpacs,aps,prd,floatfix,superscriptaddress]{revtex4}

\usepackage{graphicx}
\usepackage{dcolumn}
\usepackage{amsmath}
\usepackage{epsfig}
\usepackage{bm}
\usepackage{color}

\RequirePackage{xspace}

\usepackage{relsize}
\def\babar{\mbox{\slshape B\kern-0.1em{\smaller A}\kern-0.1em
    B\kern-0.1em{\smaller A\kern-0.2em R}}}

\def\epem       {\ensuremath{e^+e^-}\xspace}

\def\Kbar  {\kern 0.2em\overline{\kern -0.2em K}{}\xspace}

\def\Kz    {\ensuremath{K^0}\xspace}
\def\Kzb   {\ensuremath{\Kbar^0}\xspace}
\def\KzKzb {\ensuremath{\Kz \kern -0.16em \Kzb}\xspace}
\def\Kp    {\ensuremath{K^+}\xspace}
\def\Km    {\ensuremath{K^-}\xspace}

\def\KpKm  {\ensuremath{\Kp \kern -0.16em \Km}\xspace}

\def\Dbar    {\kern 0.2em\overline{\kern -0.2em D}{}\xspace}

\def\Dz      {\ensuremath{D^0}\xspace}
\def\Dzb     {\ensuremath{\Dbar^0}\xspace}
\def\DzDzb   {\ensuremath{\Dz {\kern -0.16em \Dzb}}\xspace}
\def\Dp      {\ensuremath{D^+}\xspace}
\def\Dm      {\ensuremath{D^-}\xspace}

\def\DpDm    {\ensuremath{\Dp {\kern -0.16em \Dm}}\xspace}

\def\Bbar    {\kern 0.18em\overline{\kern -0.18em B}{}\xspace}

\def\Bz      {\ensuremath{B^0}\xspace}
\def\Bzb     {\ensuremath{\Bbar^0}\xspace}
\def\BzBzb   {\ensuremath{\Bz {\kern -0.16em \Bzb}}\xspace}
\def\Bu      {\ensuremath{B^+}\xspace}
\def\Bub     {\ensuremath{B^-}\xspace}

\def\BpBm    {\ensuremath{\Bu {\kern -0.16em \Bub}}\xspace}

\def\BorBbar    {\kern 0.18em\optbar{\kern -0.18em B}{}\xspace}
\def\DorDbar    {\kern 0.18em\optbar{\kern -0.18em D}{}\xspace}
\def\KorKbar    {\kern 0.18em\optbar{\kern -0.18em K}{}\xspace}

\mathchardef\Upsilon="7107
\def\Y#1S{\ensuremath{\Upsilon{(#1S)}}\xspace}

\mathchardef\Deltares="7101
\mathchardef\Xi="7104
\mathchardef\Lambda="7103
\mathchardef\Sigma="7106
\mathchardef\Omega="710A

\def\Deltabar{\kern 0.25em\overline{\kern -0.25em \Deltares}{}\xspace}
\def\Lbar{\kern 0.2em\overline{\kern -0.2em\Lambda\kern 0.05em}\kern-0.05em{}\xspace}
\def\Sigbar{\kern 0.2em\overline{\kern -0.2em \Sigma}{}\xspace}
\def\Xibar{\kern 0.2em\overline{\kern -0.2em \Xi}{}\xspace}
\def\Obar{\kern 0.2em\overline{\kern -0.2em \Omega}{}\xspace}
\def\Nbar{\kern 0.2em\overline{\kern -0.2em N}{}\xspace}
\def\Xb{\kern 0.2em\overline{\kern -0.2em X}{}\xspace}

\newcommand{\tev}{\ensuremath{\mathrm{\,Te\kern -0.1em V}}\xspace}
\newcommand{\gev}{\ensuremath{\mathrm{\,Ge\kern -0.1em V}}\xspace}
\newcommand{\mev}{\ensuremath{\mathrm{\,Me\kern -0.1em V}}\xspace}
\newcommand{\kev}{\ensuremath{\mathrm{\,ke\kern -0.1em V}}\xspace}
\newcommand{\ev}{\ensuremath{\mathrm{\,e\kern -0.1em V}}\xspace}
\newcommand{\gevc}{\ensuremath{{\mathrm{\,Ge\kern -0.1em V\!/}c}}\xspace}
\newcommand{\mevc}{\ensuremath{{\mathrm{\,Me\kern -0.1em V\!/}c}}\xspace}
\newcommand{\gevcc}{\ensuremath{{\mathrm{\,Ge\kern -0.1em V\!/}c^2}}\xspace}
\newcommand{\mevcc}{\ensuremath{{\mathrm{\,Me\kern -0.1em V\!/}c^2}}\xspace}

\def\invfb   {\ensuremath{\mbox{\,fb}^{-1}}\xspace}

\def\mus  {\ensuremath{\rm \,\mus}\xspace}

\def\mus        {\ensuremath{\,\mu{\rm s}}\xspace}

\def\pep2{PEP-II}

\newcommand{\dedx}{\ensuremath{\mathrm{d}\hspace{-0.1em}E/\mathrm{d}x}\xspace}

\def\gsim{{~\raise.15em\hbox{$>$}\kern-.85em
          \lower.35em\hbox{$\sim$}~}\xspace}
\def\lsim{{~\raise.15em\hbox{$<$}\kern-.85em
          \lower.35em\hbox{$\sim$}~}\xspace}

\xspace

\newcommand{\jprlBase}       {Phys.\ Rev.\ Lett.\xspace}

\newcommand{\jprl}      [1]  {\jprlBase\ {\bf #1}}

\def\jetset74   {\mbox{\tt Jetset \hspace{-0.5em}7.\hspace{-0.2em}4}\xspace}

\newcommand{\kevcc}{\ensuremath{{\mathrm{\,ke\kern -0.1em V\!/}c^2}}\xspace}
\def\SC{\ensuremath{\Xi_{cc}^{+}}\xspace}
\def\DC{\ensuremath{\Xi_{cc}^{++}}\xspace}

\def\SCD{\ensuremath{\Lambda_c^+K^-\pi^+}\xspace}
\def\DCD{\ensuremath{\Lambda_c^+K^-\pi^+\pi^+}\xspace}
\def\LcD{\ensuremath{\Lambda_{c}^{+}\rightarrow pK^-\pi^+}}

\def\pcm {\ensuremath{p^*}\xspace}
\def\babar{\mbox{\slshape B\kern-0.1em{\smaller A}\kern-0.1em B\kern-0.1em{\smaller A\kern-0.2em R}}\xspace}

\def\ccd         {\ensuremath{\Xi_{cc}^+}\xspace}

\def\ccu         {\ensuremath{\Xi_{cc}^{++}\xspace}}
\def\Lc          {\ensuremath{\Lambda_c^+}\xspace}

\def\Lm          {\ensuremath{\Lambda}\xspace}
\def\CsM         {\ensuremath{\Xi^-}\xspace}
\def\CsC         {\ensuremath{\Xi_c^0}\xspace}

\def\Lmppi       {\ensuremath{\Lambda \rightarrow p \pi^-}\xspace}
\def\CsMLmpi     {\ensuremath{\Xi^-\rightarrow \Lambda \pi^-}\xspace}
\def\CsCCsMpi    {\ensuremath{\Xi_{c}^0\rightarrow \Xi^- \pi^+}\xspace}

\def\ccdCsCpi    {\ensuremath{\Xi_{cc}^+\rightarrow \Xi_c^0 \pi^+}\xspace}
\def\ccuCsCpipi  {\ensuremath{\Xi_{cc}^{++}\rightarrow \Xi_c^0 \pi^+\pi^+}\xspace}

\def\csq     {\ensuremath{c^2}\xspace}

\newcommand{\BABARPubYear}    {06}
\newcommand{\BABARPubNumber}  {031}

\newcommand{\SLACPubNumber} {11866}

\def\figurebox#1#2#3{%
    \def\arg{#3}%
    \ifx\arg\empty
    {\hfill\vbox{\hsize#2\hrule\hbox to #2{\vrule\hfill\vbox to #1{\hsize#2\vfill}\vrule}\hrule}\hfill}%
    \else
    {\hfill\epsfbox{#3}\hfill}%
    \fi}

\begin{document}

\begin{flushleft}
\babar-PUB-\BABARPubYear/\BABARPubNumber\\
SLAC-PUB-\SLACPubNumber\\
hep-ex/0605075v1
\end{flushleft}

\title{
{\large \boldmath
Search for Doubly Charmed Baryons $\Xi_{cc}^{+}$ and $\Xi_{cc}^{++}$ in \babar}
}

\author{B.~Aubert}
\author{R.~Barate}
\author{M.~Bona}
\author{D.~Boutigny}
\author{F.~Couderc}
\author{Y.~Karyotakis}
\author{J.~P.~Lees}
\author{V.~Poireau}
\author{V.~Tisserand}
\author{A.~Zghiche}
\affiliation{Laboratoire de Physique des Particules, F-74941 Annecy-le-Vieux, France }
% ...
\collaboration{The \babar\ Collaboration}
\noaffiliation

\begin{abstract}
We search for the production of doubly charmed baryons in {\ensuremath{e^+e^-}\xspace} annihilations at or near a center-of-mass energy of 10.58~GeV, in a data sample with an integrated luminosity of 232{\ensuremath{\mbox{\,fb}^{-1}}\xspace} recorded with the {\mbox{\slshape B\kern-0.1em{\smaller A}\kern-0.1emB\kern-0.1em{\smaller A\kern-0.2em R}}}\ detector at the PEP-II storage ring at the Stanford Linear Accelerator Center.
We search for $\Xi_{cc}^{+}$ baryons in the final states $\Lambda_c^+K^-\pi^+$ and $\Xi_c^0\pi^+$, and $\Xi_{cc}^{++}$ baryons in the final states $\Lambda_c^+K^-\pi^+\pi^+$ and $\Xi_c^0\pi^+\pi^+$.
We find no evidence for the production of doubly charmed baryons.
\end{abstract}


\maketitle

\setcounter{footnote}{0}

\section{Introduction and Overview}
\label{sec:Introduction}
The lowest-mass doubly charmed baryons are predicted to be the members of an isospin doublet ($\Xi^{+}_{cc}=ccd$ and $\Xi^{++}_{cc}=ccu$~\cite{conj}) 
with $J^P={1\over2}^+$ and $L=0$.
There are many theoretical predictions for the \SC and \DC masses and lifetimes~\cite{Cicerone,derujula,Fleck1,Ron,Gersh,Tong,Lewis,Itoh,Life1,Life2}.
The predicted masses lie in a range of approximately 3.5 to 3.8\gevcc~\cite{derujula,Fleck1,Ron,Gersh,Tong,Lewis}.
The mass difference between the \SC and the \DC is predicted to be on the order of 1\mevcc~\cite{Itoh}.
The \SC  and \DC lifetimes are expected to be between about 0.1 and 0.2\,ps, and 0.5 and 1.5\,ps, respectively~\cite{Life1,Life2}.
Theoretical estimates for branching fractions relevant to this paper are $B(\SC\rightarrow\SCD)=0.03$, $B(\SC\rightarrow\Xi_c^0\pi^+)=0.02$, $B(\DC\rightarrow\DCD)=0.05$, and $B(\DC\rightarrow\Xi_c^0\pi^+\pi^+)=0.05$~\cite{Bjorken}.

Several predictions have been made for the production cross sections of doubly charmed baryons in \epem annihilations~\cite{DiqProduction,Production,DiqSpinTheor};
the predictions range from 1 to 250~fb for an \epem center-of-mass (CM) energy near 10.58 GeV, and translate into $\cal{O}$$(10^2$--$10^4)$ doubly charmed baryons produced in the \babar\ data set of 232\invfb analyzed here.
Measured cross sections for double-$c\bar{c}$ production in Belle~\cite{BelleCharm} and \babar~\cite{BaBarDCharmo} are an order of magnitude larger than non-relativistic QCD predictions.
Calculations for $c\bar{c}\,c\bar{c}$ and $cc\,\bar{c}\bar{c}$ cross sections are very similar;
therefore, the predicted cross sections for doubly charmed baryons may also have been underestimated.

The SELEX collaboration, which uses the Fermilab 600-GeV/$c$ charged hyperon beam, has published evidence for the \SC baryon in the $\Lambda_c^+K^-\pi^+$ and $pD^+K^-$ decay modes with a mass of $(3518.7\pm1.7)$\,\mevcc~\cite{SELEX1,SELEX2}.
The \DC baryon, detected in the decay mode $\Lambda_c^+K^-\pi^+\pi^+$, with a mass of 3460\mevcc, was reported by SELEX at ICHEP 2002~\cite{SELEX3}.
The \SC-\DC mass difference of 60\mevcc is not consistent with theoretical expectations.
SELEX sets an upper limit (at 90\% confidence level) of 33\,fs on the lifetime of the \SC baryon, in conflict with theoretical predictions.
The photoproduction experiment FOCUS does not observe any $\Xi_{cc}$ states~\cite{FOC} although they observe $19,500$ \Lc baryons,
compared to $1,650$ for SELEX.

In this paper, we describe a search for the production of $\Xi_{cc}$ baryons in a data sample corresponding to an integrated luminosity of 232~\invfb recorded with the \babar\ detector at the PEP-II asymmetric-energy \epem storage ring at the Stanford Linear Accelerator Center.
Events containing $\LcD$ candidates are searched for the presence of $\SC\rightarrow\SCD$ and $\DC\rightarrow\DCD$ candidates.
Events containing $\Lambda\rightarrow p\pi^-$ candidates are searched for the presence of $\SC\rightarrow\Xi_c^0\pi^+$ and $\DC\rightarrow\Xi_c^0\pi^+\pi^+$ candidates where $\Xi_c^0\rightarrow\Xi^-\pi^+$ and $\Xi^-\rightarrow\Lambda\pi^-$.

The \babar\ detector is described in detail elsewhere~\cite{babar}.
The tracking of charged particles is provided by a five-layer double-sided silicon vertex tracker (SVT) and a 40-layer drift chamber (DCH).
Discrimination among charged pions, kaons, and protons relies on ionization energy loss (\dedx) in the DCH and SVT, and on Cherenkov photons detected in a ring-imaging detector (DIRC).
A CsI(Tl) crystal calorimeter is used to identify electrons and photons.
These four detector subsystems are mounted inside a 1.5-T solenoidal superconducting magnet.
The instrumented flux return for the solenoidal magnet provides muon identification.

For event simulations, we use the Monte Carlo (MC) generators JETSET74 \cite{jetset} and EVTGEN~\cite{EVTGEN} with a full detector simulation based on GEANT4~\cite{MC}.
These simulations are used to estimate the reconstruction efficiencies of the searches.
For each of the four $\Xi_{cc}$ decay channels used in our searches, we produce approximately 100,000 simulated $\epem\rightarrow c\bar{c}$ events in which at least one of the primary charm quarks hadronizes into a $\Xi_{cc}$.
The distribution of momentum in the CM frame (\pcm) for simulated $\Xi_{cc}$ peaks at about 2.5\gevc, with 80\% above 2.0\gevc and 62\% above 2.3\gevc.
The \SC and \DC baryons are simulated with the SELEX masses of 3520 and 3460\mevcc, respectively.
The \SC, \DC, and $\Lambda_c^+$ decays are generated according to phase space.

We search for $\Xi_{cc}$ production as an excess of candidates in the
distribution of the difference in the measured masses of the $\Xi_{cc}$ and the candidate daughter baryon.
Some mass uncertainties cancel in this mass difference, 
improving the mass resolution and thereby the
signal-to-background ratio.
We use the following notation: $\Delta M(A -B) \equiv M(A) - M(B)$, where $A$ is the parent and $B$ is the daughter baryon.
$M(X)$ refers to the measured invariant mass of the $X$ candidate.

Selection criteria are chosen to maximize $\epsilon/\sqrt{B}$, where $\epsilon$ is the simulated reconstruction efficiency and $B$ is the number of candidates in data in upper and lower sidebands of the mass-difference regions in which we search for $\Xi_{cc}$ signals.
During this process the search regions were hidden to minimize potential experimenter bias.

Charm hadrons carry a significant fraction of the initial energy of the charm quark, whereas random combinations of charged particles in an event form lower-energy candidates.
To take advantage of this difference,
we select $\Xi_{cc}$ candidates for which the \pcm of the $\Xi_{cc}$ is above a minimum value.  
For $\Xi_{cc}$ decay modes containing a \Lc, the optimal requirement is $\pcm > 2.3\gevc$.  
Because the background levels for events containing a $\Xi_c$ candidate are lower, we apply the less stringent requirement $\pcm > 2.0\gevc$.
To facilitate comparisons with theoretical predictions, we repeat the searches with no requirement on \pcm.

We conduct searches for $\Xi_{cc}$ near the masses of the states observed by SELEX and over wider ranges that include many of the theoretically predicted masses.
We use MC techniques to account for the width of the search region in the statistical interpretation of the results.

\section{\boldmath Search for Decays to $\Lambda_c^+ K^- \pi^+ (\pi^+)$}
\label{sec:lambdac}
In the searches for $\SC\rightarrow\Lambda_c^+K^-\pi^+$ and $\DC\rightarrow\Lambda_c^+K^-\pi^+\pi^+$, we reconstruct the \Lc baryon in its decay to $pK^-\pi^+$.
Pion, kaon and proton candidates are identified using the 
SVT, DCH and DIRC.
The $\chi^2$ probability for the \Lc daughter particles and for the $\Xi_{cc}$ daughter particles to each come from a common vertex is required to be above 1\%.
The number of reconstructed \Lc signal events is approximately $600,000$.

\begin{figure}[!tbh]
\includegraphics[width=0.5\textwidth]{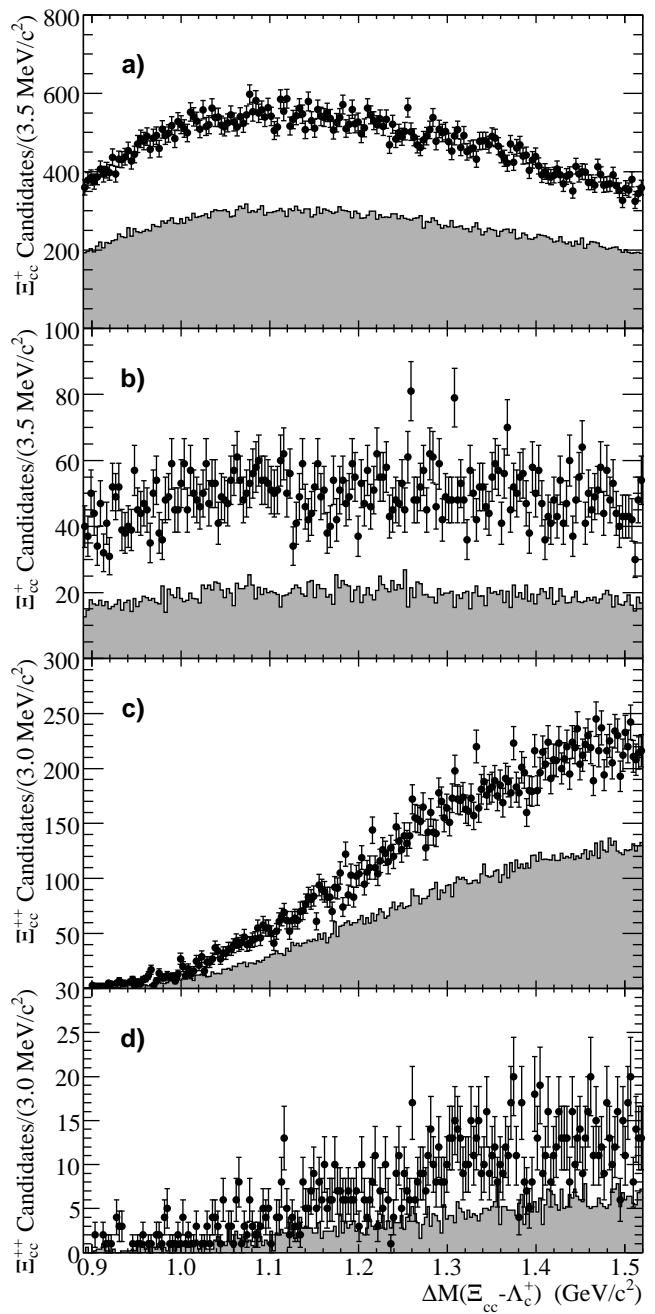}
\caption{Distributions of the mass difference $\Delta M(\Xi_{cc} - \Lc)$ for (a,b) \ccd and (c,d) \ccu candidates with (a,c) no \pcm requirement and (b,d) \pcm $>2.3$~GeV/$c$.
Data points with error bars correspond to candidates near the \Lc mass: 
$2281<M(\Lc)<2291\mevcc$.
Shaded histograms correspond to candidates in $M(\Lc)$ sidebands ($2256<M(\Lc)<2281$\mevcc and $2291<M(\Lc)<2316$\mevcc), scaled to represent the expected amount of non-\Lc background in the data projections.}
\label{fg:lcdata}
\end{figure}

The distribution of the mass difference $\Delta M(\Xi_{cc}-\Lambda_c^+)$ is shown in Fig.~\ref{fg:lcdata} for candidates with $M(\Lc)$ between 2281 and 2291\mevcc ($\pm0.8\sigma$), and also for $M(\Lc)$ sidebands ($2256<M(\Lc)<2281$\mevcc and $2291<M(\Lc)<2316$\mevcc).
To search for a signal in data and to estimate the efficiency, we perform two-dimensional fits to $M(\Lambda_c^+)$ and $\Delta M(\Xi_{cc}-\Lambda_c^+)$.
The range of $M(\Lambda_c^+)$ used in all fits is 2256 to 2316\mevcc.
We search for $\Xi_{cc}$ states with masses between 3390 and 3600\mevcc
($\Delta M(\Xi_{cc}-\Lc)$ between 1100 and 1310\mevcc).
The mass-difference sidebands in data are between 890 and 1100\mevcc, and
1310 and 1520\mevcc.

Approximately half of all background $\Xi_{cc}$ candidates are due to true \Lc particles combined with random pion and kaon candidates from the rest of the event.
This background is fit with a Gaussian shape in $M(\Lc)$ and a linear shape in $\Delta M(\Xi_{cc}-\Lc)$.
Another significant background contribution is from false \Lc candidates.
This source of background is fit with the product of a linear function in $M(\Lc)$ and a linear function in $\Delta M(\Xi_{cc}-\Lc)$.

MC simulations show that $\Xi_{cc}$ signals peak in three different ways in the $M(\Lc)$ versus $\Delta M(\Xi_{cc}-\Lc)$ plane.
In most cases, the $\Xi_{cc}$ is reconstructed correctly and the measured values of both $M(\Lc)$ and $\Delta M(\Xi_{cc}-\Lc)$ lie close to the generated values; 
such candidates are fit with the product of two Gaussian distributions, one in each variable.
The MC signal resolution for $\Delta M(\Xi_{cc}-\Lc)$ is 3.5\mevcc and 3.0\mevcc for \SC and \DC, respectively.
When $\Xi_{cc}$ candidates are reconstructed from the correct tracks but the kaon and/or pion from the \Lc decay is swapped with the kaon and/or pion from the $\Xi_{cc}$ decay, the reconstruction has the correct $M(\Xi_{cc})$ but an incorrect $M(\Lc)$.
These events are fit in both MC simulations and data with a 
Gaussian function in $\Delta M(\Xi_{cc}-\Lc)+M(\Lc)= M(\Xi_{cc})$ and are included as part of the signal.
When the \Lc is correctly reconstructed but is combined with an incorrect pion and/or kaon to form the $\Xi_{cc}$, the reconstruction has the correct $M(\Lc)$ but an incorrect $\Delta M(\Xi_{cc}-\Lc)$.
Such events are not distinguishable from \Lc combinatoric background.

\begin{figure}[!tbh]
\includegraphics[width=0.5\textwidth]{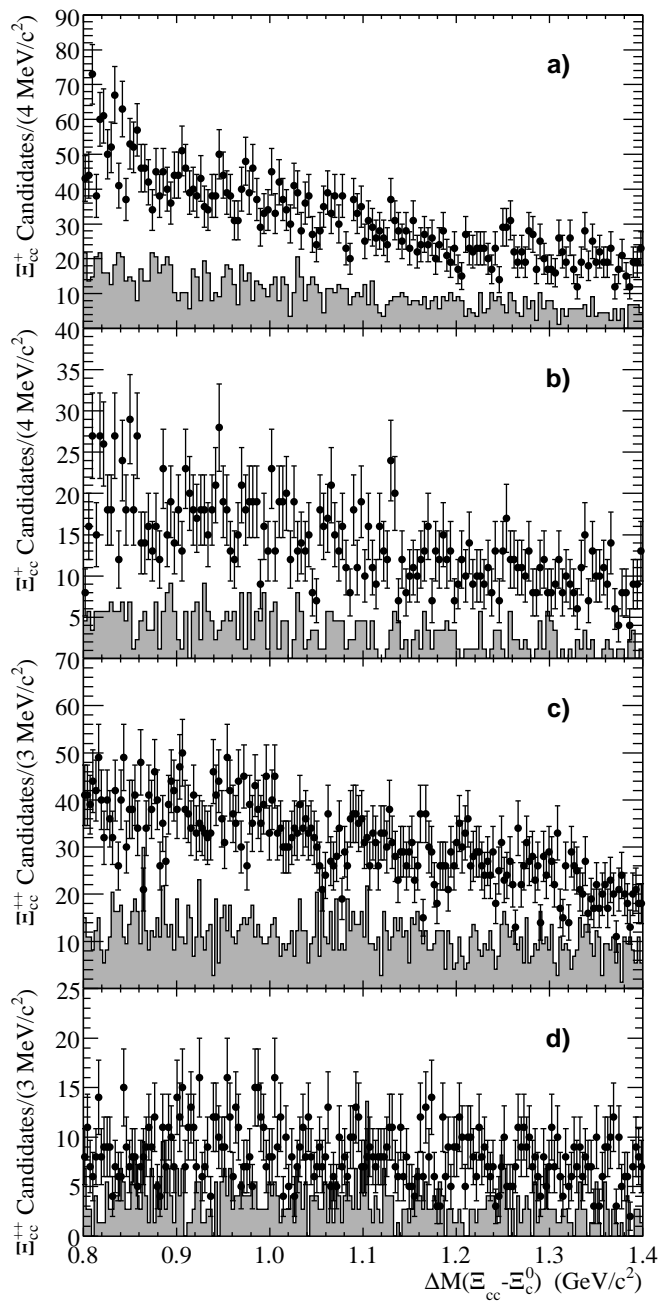}
\caption{Distributions of the mass difference $\Delta M(\Xi_{cc} - \CsC)$ for (a,b) \ccd and (c,d) \ccu candidates with (a,c) no \pcm requirement and (b,d) \pcm $>2.0$~GeV/$c$. Data points with error bars correspond to $\Xi_{cc}$ candidates reconstructed using \CsC candidates near the $\CsC$ mass, $2457<M(\CsC)<2485$\mevcc; the shaded histograms correspond to $M(\CsC)$ sidebands ($2451<M(\CsC)<2457$\mevcc and $2487<M(\CsC)<2501$\mevcc) scaled to represent the expected amount of non-\CsC background in the $M(\CsC)$ signal region.
\label{fig:Ycc_all}}
\end{figure}

Each shape parameter describing the signal is constrained in the fit to lie within a range determined from the Monte Carlo simulation, allowing for possible inaccuracies in the simulation.
The integral of the signal function is allowed to be negative.
Efficiencies for the reconstruction of $\Xi_{cc}$ baryons decaying to $\Lc K^-\pi^+$ and $\Lc K^-\pi^+\pi^+$ are calculated from the signal yields from fits to the MC simulated samples.
These efficiencies are listed in Table~\ref{tb:effcalc}.
The systematic uncertainties are due to inaccuracies in the simulation of tracking reconstruction (0.8\% per track, added linearly) and particle identification (1.0\% per kaon, 1.0\% per pion, and 4.0\% per proton).
When setting upper limits on production cross sections, additional systematic uncertainties arise due to the uncertainties on the integrated luminosity (1.0\%) and $\sigma(\epem\rightarrow\Lc X)B(\LcD)$ (4.7\%).

\begin{table}[!tbh]
\caption{\label{tb:effcalc} 
Efficiencies determined from $\epem\rightarrow\Xi_{cc}X$ simulations.
With the $p^*$ criterion applied, the efficiency is calculated for $\Xi_{cc}$
baryons generated with $p^*$ above 
2.3~GeV/$c$ for the \Lc modes and 2.0~GeV/$c$ for the $\Xi_c$ modes.
The first error is statistical; the second is systematic.}
\begin{center}
\begin{tabular}{cccc}\hline\hline
$p^*$ Criterion & Particle    & \Lc Mode Eff. (\%) & $\Xi_c^0$ Mode Eff. (\%) \\ \hline
Yes             & \DC         & $ 4.2\pm0.1\pm0.2$ & $6.1\pm0.1\pm0.6$        \\
Yes             & \SC\        & $10.4\pm0.1\pm0.5$ & $9.3\pm0.1\pm0.7$        \\
No              & \DC         & $ 3.6\pm0.1\pm0.2$ & $5.9\pm0.1\pm0.5$        \\ 
No              & \SC\        & $ 9.7\pm0.1\pm0.5$ & $9.0\pm0.1\pm0.7$        \\ \hline\hline
\end{tabular}
\end{center}
\end{table}

We conduct searches for a signal within 10-MeV/$c^2$-wide regions around the \SC and \DC masses reported by SELEX, and within the 210-MeV/$c^2$-wide region described earlier.
The wide search region is divided into 21 sequential 10-MeV/$c^2$ search sub-regions.
For each sub-region, we perform a two-dimensional fit over a 100-MeV/$c^2$-wide range in mass difference centered on the sub-region, constraining the mean of the Gaussian signal function to lie within that sub-region.

The significance of any potential signal is determined through the use of parametrized MC simulations.
Samples of pairs of variables ($M(\Lc)$, $\Delta M(\Xi_{cc}-\Lc)$) are generated according to the background shapes measured in data, with no signal contribution.
The distributions of $M(\Lc)$ versus $\Delta M(\Xi_{cc}-\Lc)$ from these simulations are then searched in the same manner as in data.
A significance measure $N/\sigma_N$, where $N$ is the fitted number of signal candidates and $\sigma_N$ is the uncertainty on this number, is determined for each fit.
In order to statistically combine the results of the 21 fits into one search, only the largest of the 21 significance measures is used.
The significance measure from data is compared to the distribution of significance measures from the MC simulations that represent those data.
This comparison gives the probability of measuring this particular value of $N/\sigma_N$ or higher in data under the hypothesis that no $\Xi_{cc}$ are produced.

None of the \Lc decay mode searches finds evidence for $\Xi_{cc}$.
The most statistically significant signal is for a \SC baryon with $\Delta M(\Xi_{cc}^{+}-\Lc)$ between 1250\mevcc and 1260\mevcc, when candidates are required to have $\pcm>2.3\gevc$.
With a significance measure of $N/\sigma_N = 66/24$, 
we find that there is an 8\% probability that background alone could produce this signal.
This corresponds to a significance of $1.4\,\sigma$, which does not constitute evidence for the \SC baryon.

Using efficiencies ($\epsilon$) listed in Table~\ref{tb:effcalc} and integrated luminosity ($L$) of $(232\pm2)$\,\invfb, we extract values for the upper limit on the production cross section times branching fraction(s) (${\cal S}$) directly from negative-log-likelihood functions.
A conversion factor ${\cal F}=L\epsilon$ and its uncertainty $\sigma_{\cal F}$ are incorporated in a Gaussian extension to the likelihood function (${\cal L}$) so that all systematic uncertainties are included in the results.
${\cal L}$ takes the form
\begin{equation*}{\cal L}=e^{-(N-{\cal S}f-n_b)} e^{-\frac{({\cal F}-f)^{2}}{2\sigma_{\cal F}^2}} \prod_{i}^{N}P(\vec{x}_i;{\cal S},f,n_b,\vec{a})\,,\end{equation*}
where $N$ is the total number of fitted events; ${\cal S}f=n_s$ and $n_b$ are the fitted number of signal and background events, respectively; $f$ is the fitted conversion factor from ${\cal S}$ to $n_s$; $\vec{a}$ are shape parameters; and $P$ is the probability function for the data point $\vec{x}_i$.
The value of ${\cal S}$ for which $-\ln{\cal L}$ is 1.35 units above the minimum value for which ${\cal S}$ is positive is interpreted as the 95\%-confidence-level upper limit.
These limits are listed in Table~\ref{tab:physics_Lc_UL}.

\begin{table*}[!tbh]
\caption{
The 95\%-confidence-level upper limits on measured rates for the production of $\Xi_{cc}$ baryons with and without a $p^*$ requirement of 2.3~GeV/$c$ for \Lc modes and 2.0~GeV/$c$ for the $\Xi_c$ modes.
The columns labeled $N^{+(+)}$ give the upper limits on the number of signal $\Xi_{cc}^{+(+)}$ baryons.
$\sigma^{+(+)}$ denotes the production cross section $\sigma(\epem\rightarrow\Xi_{cc}^{+(+)}X)$;
$\sigma$ in the denominator indicates that the cross section has been normalized to $\sigma(\epem\rightarrow\Lc X)B(\LcD)$.
The factor $B$ in a column heading signifies that the values in the column correspond to a cross section times the branching fractions 
$B(\Xi_{cc}^{+(+)}\rightarrow\Lc K^-\pi^+(\pi^+))B(\LcD)$
for decay modes with \Lc and  
$B(\Xi_{cc}^{+(+)}\rightarrow\Xi_c^0\pi^+(\pi^+))B(\CsCCsMpi)$
for decay modes with $\Xi_c^0$.
For each wide mass range, the upper limit corresponds to the maximum upper limit over the range.}
\begin{center}
\begin{tabular}{cccccccccccccc} \hline \hline
& \multicolumn{7}{c}{Upper Limits for $\Xi_{cc}\rightarrow\Lc K^{-}\pi^{+}(\pi^{+})$} &~~~~~& \multicolumn{5}{c}{Upper Limits for $\Xi_{cc}\rightarrow\Xi_c^0\pi^+(\pi^+)$} \\
&$N^{+}$&$\sigma^{+}B$&$(\sigma^{+}/\sigma)B$&~~~&$N^{++}$&$\sigma^{++}B$&$(\sigma^{++}/\sigma)B$&~~~~~~~~&~~~$N^{+}$~~~&~$\sigma^{+}B$~&~~~~~&$N^{++}$&$\sigma^{++}B$\\ \hline
Wide Mass Range~~~~~
& 328   & 14.5 fb     & $13.3\times10^{-4}$  &~~~& 199    & 23.9 fb      & $22.0\times10^{-4}$   &~~~~~~~~& 58    & 4.3 fb      &~~~& 58     & 6.6 fb       \\

Wide Mass Range, \pcm Req.~~~~~
& 106   &  4.4 fb     &  $5.6\times10^{-4}$  &~~~& 54     &  5.5 fb      &  $6.9\times10^{-4}$   &~~~~~~~~& 41    & 3.0 fb      &~~~& 28     & 3.1 fb       \\   

SELEX Mass~~~~~
& 169   &  7.5 fb     &  $6.9\times10^{-4}$  &~~~& 91     & 10.9 fb      &  $10.0\times10^{-4}$   &~~~~~~~~& 26    & 2.0 fb      &~~~& 49     & 5.6 fb       \\

SELEX Mass, \pcm Req.~~~~~
& 53    &  2.2 fb     &  $2.7\times10^{-4}$  &~~~& 31     &  3.2 fb      &  $4.0\times10^{-4}$   &~~~~~~~~& 18    & 1.3 fb      &~~~& 31     & 3.4 fb       \\ 

\hline \hline
\end{tabular}
\end{center}
\label{tab:physics_Lc_UL}
\end{table*}

To facilitate comparison with the production rate of \Lc and to take advantage of the cancellation of the \LcD\ branching fraction, we also normalize the upper limits to $\sigma(\epem\rightarrow\Lc X)B(\LcD)$, measured with 22\invfb of data collected at $\sqrt{s}\sim 10.54\gev$;
these upper limits are also listed in Table~\ref{tab:physics_Lc_UL}.
The \pcm criterion that is applied to the $\Xi_{cc}$ candidates is also applied to the \Lc candidates in the normalization mode.

\section{\boldmath Search for Decays to $\Xi_c^0 \pi^+ (\pi^+)$}
\label{sec:xic}
In the search for \ccdCsCpi and \ccuCsCpipi decays, the \CsC is
detected in the decay chain \CsCCsMpi, \CsMLmpi, $\Lambda
\rightarrow p\pi^-$. 
We search for $\Xi_{cc}$ states with 
masses between 3370 and 3770\mevcc ($\Delta M(\Xi_{cc} -
\Xi_{c}^0)$ between 900 and 1300\mevcc). 
The mass-difference sidebands in data are
$800<\Delta M(\Xi_{cc} - \Xi_{c}^0)<900\mevcc$ and 
$1300<\Delta M(\Xi_{cc} - \Xi_{c}^0)<1400\mevcc$.

For \Lm and \CsM candidates, we require a minimum signed
three-dimensional flight distance of $+2.0$\,cm and $+0.5$\,cm,
respectively, where the flight distance is the projection of the vector from
the primary vertex to the decay point, onto the momemtum
vector of the candidate. \Lm candidates are required to be within $\pm
3.6\mevcc$ ($\pm 3\sigma$) of the world average mass~\cite{RPP}. \CsM candidates
are required to be within $\pm 5.4\mevcc$ ($\pm 3\sigma$) of the world
average mass difference $\Delta M(\CsM-\Lm)$, and \CsC candidates are
required to be within $\pm 14\mevcc$ ($\pm 2\sigma$) of the world
average mass difference $\Delta M(\CsC-\CsM)$~\cite{RPP}. For all candidate
baryons, we require the vertex fit to have a $\chi^2$ probability greater
than 0.01\%. 
The number of reconstructed \CsC signal events is approximately $11,700$.
Figure~\ref{fig:Ycc_all} shows the distributions of mass
difference for all $\Xi_{cc}$ candidates that satisfy these criteria,
with no \pcm requirement and with \pcm$>2.0$~GeV/$c$. The
reconstruction efficiencies are given in Table~\ref{tb:effcalc}.

Systematic uncertainties arise mainly from possible inaccuracies in
the simulation of 
track reconstruction and particle identification (5\% for \ccd
and 6\% for \ccu), vertex quality (6\%), and mass and mass-difference
resolutions (1\%); the values in parentheses are the relative
uncertainties in these efficiencies.  Other sources include
uncertainties in the total luminosity (1.0\%) and in the branching
fractions for \Lmppi (0.8\%) and \CsMLmpi (0.03\%).

To search for a signal in the 400-MeV/\csq-wide search region, 
we fit the mass-difference distribution with two Gaussian functions,
with common means and fixed widths, to represent the signal, and a
first-order polynomial for the background. The values of the Gaussian
widths are determined from the MC simulation; 
the root-mean-squared deviation for 
$\Delta M(\ccd - \CsC)$ is 5.5~MeV/\csq and for
$\Delta M(\ccu - \CsC)$ it is 4.2~MeV/\csq.  We conduct 50 fits with
the mean of the Gaussian signal function constrained to lie in 50
10-MeV/\csq ranges, each of which overlaps neighboring ranges by
2~MeV/\csq.  
Using a MC approach, we calculate the upper limit on the number of 
signal events using the statistically most significant of the 50 fits.
To do this, we generate
$N$ signal events according to the Gaussian signal function 
and background events 
according to a first-order polynomial,
where the number of background events is determined from the mass-difference sidebands. 
We fit the resulting MC distribution as described above for data,
and record the number of signal events $S$ for 
the statistically most significant fit. 
We repeat this process 10,000 times, 
varying $N$ by the fractional systematic uncertainty on efficiency.
We then find the value $F$ for
which only 5\% of the trials have $S<F$.
We repeat the above process starting with different values of $N$ to
find the value of $N$ for which $F$ is the number of signal events
found in the most significant fit in data.
This value of $N$ is the 95\% CL upper limit on the number of events,
shown in Table~\ref{tab:physics_Lc_UL}
for both \ccd and \ccu, with and without \pcm requirements. 
We also present in Table~\ref{tab:physics_Lc_UL} the limits obtained
when we explicitly search for the states observed by SELEX. 
For comparison, the measured rate for the singly charmed 
$\Xi_c$ baryon in \babar is 
$\sigma(\epem\rightarrow\CsC X)B(\CsCCsMpi) = (388\pm39\pm41)$~fb~\cite{Casc_cross}.

\section{Summary}
In conclusion, we have searched for doubly charmed baryons in \epem annihilations at or near a center-of-mass energy of 10.58\gev.
We do not observe any significant signals for
the \SC baryon in the decay modes $\Lambda_c^+K^-\pi^+$ and $\Xi_c^0\pi^+$, or for the \DC baryon in the decay modes $\Lambda_c^+K^-\pi^+\pi^+$ and $\Xi_c^0\pi^+\pi^+$.

\vskip0.3cm
We are grateful for the excellent luminosity and machine conditions
provided by our \pep2\ colleagues, 
and for the substantial dedicated effort from
the computing organizations that support \babar.
The collaborating institutions wish to thank 
SLAC for its support and kind hospitality. 
This work is supported by
DOE
and NSF (USA),
NSERC (Canada),
IHEP (China),
CEA and
CNRS-IN2P3
(France),
BMBF and DFG
(Germany),
INFN (Italy),
FOM (The Netherlands),
NFR (Norway),
MIST (Russia), and
PPARC (United Kingdom). 
Individuals have received support from CONACyT (Mexico), 
Marie Curie EIF (European Union),
the A.~P.~Sloan Foundation, 
the Research Corporation,
and the Alexander von Humboldt Foundation.


\begin{thebibliography}{99}
\bibitem{conj} Throughout this paper, whenever a particle or decay mode 
is given, the charge conjugate is also implied.
\end{thebibliography}
\end{document}